\documentclass[twoside]{article}
\usepackage{eadh2026}
\usepackage{longtable}
\usepackage{booktabs}       
\usepackage{microtype}      
\usepackage{ltxtable}
\usepackage{graphicx}
\bibliography{eadh2026.bib}
\begin{document}

\title{Linking Speakers of the German Parliament to Wikidata: Scope and Coverage of Metadata}

\author[1]{Thomas Haider \orcid{0000-0003-1522-4026}}
\author[1]{Arne Cypionka \orcid{0009-0001-4984-1631}}
\author[1]{Maximilian Teich \orcid{0009-0000-7084-8291}}

\affil[1]{Chair of Computational Humanities, University of Passau}

\titlerunning{Linking Speakers of the German parliament to Wikidata} 
\authorrunning{Haider et al.} 

\maketitle   
\thispagestyle{empty}


\begin{abstract}
This paper links all individuals who spoke in the German Bundestag between 1949 and 2021 to Wikidata, creating a longitudinal
dataset that connects parliamentary speech transcripts with structured biographical metadata. We evaluate the coverage, composition, and potential biases of the retrieved properties and statements, with particular attention to gender, professional background, historical legacies, and transnational dimensions such as place of birth, languages, and foreign awards. The results demonstrate both the analytical potential of combining GermaParl 
with Wikidata and the importance of critically assessing uneven metadata coverage in open knowledge graphs.
\end{abstract}


\section{Introduction}
The German federal parliament (Bundestag) publishes its plenary proceedings together with a directory of all elected members (so called `Stammdaten').\footnote{\url{https://www.bundestag.de/services/opendata}} This directory provides core biographical metadata, consisting of names, gender, and party affiliation during a member's term of office. 
While the records remain limited in their breadth and depth, in combination with the speech transcripts, this information already enables considerable analysis. As studies on populism \parencite{erhard_popbert._2025}, gender roles \parencite{Spieker2021, stecker_evolution_2021}, or the political left–right spectrum \parencite{warode_mapping_2025} demonstrate, there is considerable analytical potential and sustained scholarly interest in utilizing the Bundestag plenary proceedings with the metadata available. 

At the same time, the official member directory (`Stammdaten') only covers elected representatives. The plenary proceedings, however, include a broader set of speakers, such as members of the federal government, parliamentary state secretaries, and invited guests. In total, over 4,000 people spoke in front of Germany's parliament.  As can be seen in Table \ref{tab:properties} only 87.5\% of these speakers are elected officials (P39 - member of the German Bundestag). For the other individuals, no consolidated biographical metadata are systematically provided alongside the proceedings. 

Further information about these persons is encoded in various resources, such as DBpedia \parencite[cf.][]{leonhardt-blaette-2024-dbpedia} and consolidated metadata databases. For instance, \textcite{gobel_comparative_2022} and \textcite{herrmann_party_2023} have worked on expanding and refining available metadata for parliamentary actors, showing that enriched contextual information substantially broadens the range of possible inquiries. These efforts, however, remain confined 
to the databases themselves
and, to our knowledge, do not yet comprehensively map the parliamentary speech data to the metadata. These databases can be accessed through Wikidata, to which this paper provides the linking. 





Especially the historical dimension of this data make it paramount to have open resources that can
be critically examined. 
Social and ideological biases in parliaments are well documented.
\textcite{navarretta2022subject}, for instance, find gendered patterns in the Danish parliament, where women disproportionately engage with social issues and men with economic ones. \textcite{schulz_braune_2021} in turn, aims to trace how National Socialist ideology persisted into the parliaments of the early Federal Republic. 
However, to our knowledge, a systematic evaluation of coverage and  potential data biases has so far been lacking, both in the parliamentary source data and in the linked external resources. 
Here we address this gap.

The contribution of this paper is to link every person to Wikidata who spoke in the German Bundestag from 1949 to 2021 (1st to 19th legislative period). On this basis, we assess the scope, composition, and coverage of additional metadata that can be retrieved through this linkage. To match the conference theme `Linking Europe', we also investigate place of birth, languages spoken, and awards given by foreign entities/countries. We hope to give some pointers at research directions to use this data to e.g., correlate the language of the speeches with prosopographical information.

The basis of the project consists of two primary data sources: GermaParl and Wikidata.
For the Bundestag proceedings we rely on the GermaParl corpus \parencite{blatte2018germaparl,blatte2022germaparl}, which provides a machine-readable (TEI P5) and annotated version of the plenary protocols (featuring i.a., speaker turns, interjections and party affiliation from the Stammdaten directory). GermaParl is currently being integrated into the larger ParlaMINT corpus \parencite{Erjavec2025-xv}. As part of this migration, the GermaParl team is drawing on Wikidata for person metadata -- in particular for non-MP speakers' sex and identifiers \parencite{leonhardt-blaette-2026-lrec} -- which is narrower than the coverage pursued here.

On the other hand, Wikidata serves 
as an open, collaboratively curated knowledge base containing structured biographical and contextual information on public figures, including politicians, and references to other authoritative resources like the GND (the Integrated Authority File; for which coverage is almost exhaustive as can be seen below), making it feasible to then further map to the resources of \textcite{gobel_comparative_2022} and \textcite{leonhardt-blaette-2024-dbpedia,leonhardt-blaette-2026-lrec} down the line.

\section{Matching and Metadata Extraction}

From the GermaParl corpus, we extract all annotated speakers and their associated party affiliations. In total, the dataset contains 4,325 unique speakers. The dataset is available through the Open Science Framework.\footnote{\url{https://osf.io/5s3nc/}}

To assign unique Wikidata identifiers to all recorded speakers, we implement a four-step matching procedure:
\begin{enumerate}
    \item \textbf{Name and party match:}
            For each speaker, Wikidata is queried for individuals identified as members of the Bundestag with the corresponding name and party affiliation.
    \item \textbf{Name-only match (restricted to Bundestag members):}
            If no match is found, we query for members of the Bundestag with the same name, irrespective of party affiliation.
    \item \textbf{General name match:}
            If this step also yields no result, Wikidata is queried for any person with the given name.
    \item In cases where multiple matches are returned, the correct identifier is determined manually. Manual verification is likewise conducted when no automatic match can be identified.
\end{enumerate}


Once identifiers are established, we retrieve all available Wikidata predicates and then statements (predicate-object) associated with each speaker through the Wikidata query service. Table \ref{tab:properties} shows the properties with at least 1\% coverage and Table \ref{tab:statements} the statements with at least 1\% coverage (see below in the appendix). 
These tables already give us a general overview and then we can derive specific dimensions of interest through filtering. 

\section{Results}

\subsection{Properties}

The coverage of properties (Table \ref{tab:properties}) shows that Wikidata is almost complete for the (4,325) Bundestag/parliamentary speakers in terms of their `core profile': Basic characteristics such as gender, instance of, date of birth, languages, occupation, and key political information (party, offices/positions) are predominantly at $\sim$98–100\%, and authority IDs (especially GND/VIAF) are also very strongly represented – which suggests that Wikidata is curated through these resources. 

In contrast, CV and provenance fields such as `educated at' ($\sim$24\%) or the official Bundestag biography link ($\sim$44\%) are significantly weaker, which means that the data collection is optimized for identity/roles rather than complete CVs; media and Commons data (image/Commons category) are only around half complete. 

The historical nature of the population is also important: at least 51\% of the people recorded already have a date of death (P570) and are therefore deceased, and 348 people have a NSDAP membership number (P2298), which indicates substantial personal overlap with the period 1925–1945, which now opens this data for research questions about influences of the Nazi Regime in the Bundestag \parencite{schulz_braune_2021}.

\newpage

\subsection{Statements}

\begin{figure}[htbp!]
    \centering
    \includegraphics[width=0.55\textwidth]{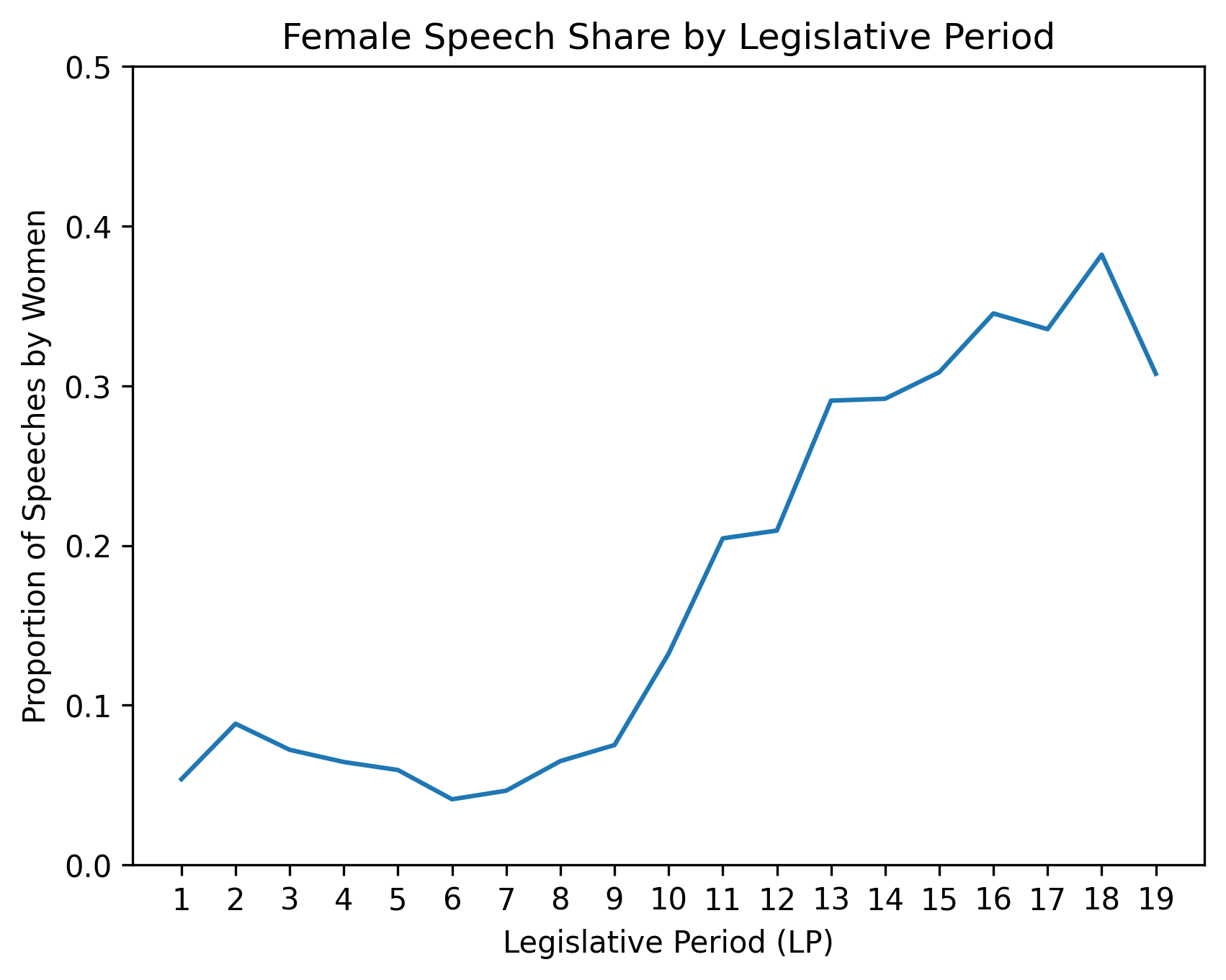}
    \caption{Proportion of speeches given by women across legislative periods (LP 1--19).}
    \label{fig:female_speech_share}
\end{figure}

The coverage of statements (Table \ref{tab:statements}) shows that gender is systematically recorded; the historical skew ($\sim$79.4\% male, $\sim$20.6\% female) reflects male-dominated early parliaments. Looking at the speeches, we find that women’s share of speeches rises from 5\% in the first legislative period to nearly 40\% in the 18th (Figure \ref{fig:female_speech_share}).

Work location data capture the institutional shift from Bonn ($\sim$64.6\%) to Berlin ($\sim$21.4\%) after reunification, though Berlin appears under-recorded; additional sparse locations likely reflect prior state-level careers. Political affiliation appears both as party membership (P102: SPD 34.0\%, CDU 32.7\%, FDP 11.8\%, CSU 7.5\%, Greens 5.9\%, Left 3.6\%, AfD 2.1\%) and parliamentary group (P1416: CDU 34.5\%, SPD 29.4\%, FDP 9.4\%).

Doctorates (4.4\%) slightly exceed general population levels (~2\% with historical fluctuations), indicating modest overrepresentation without academic elitism.

Legal and political professions dominate occupations (P106), with lawyers, economists, teachers, and civil servants prevalent. The increasing use of “politician” over time signals professionalization. While P106 records formal job titles, field of work (P101) captures broader topics/policy areas and is sparsely populated.

Finally, entries such as Nazi Party membership (369; $\sim$8.5\%),\footnote{Distinct from NSDAP membership number seen above.} SA/SS affiliations, World War participation, and citizenships like the GDR or German Reich indicate the complex historical trajectories of these political figures.

\subsection{Relationship to Europe}

Place of birth (Table \ref{tab:placeofbirth}) is dominated by major German cities but includes now Polish cities such as Wrocław (29), Szczecin (12), Gdańsk (9), and Katowice (6). Their presence in the dataset likely reflects the forced migration of Germans from Silesia and other eastern territories after WWII, showing how geopolitical border shifts are encoded in this biographical data.

Language data (Table 4) is limited: around 15\% speakers do not have German as their native language. English (as a language that is expected to be widely spoken) appears in only~1\%, likely due to incomplete Wikidata recording. Other languages (Turkish, Polish, Russian, French) occur in small shares but are probably under-reported as well.

Awards are predominantly German but include a notable number of Austrian distinctions and some from Finland and Italy, indicating recognition beyond the national level that reflects German–Austrian ties as well as German war-time alliances.

\section{Conclusion}

In this paper, we demonstrate that systematically linking all Bundestag speakers from 1949 to 2021 to Wikidata yields high coverage for core biographical and political attributes, while revealing clear gaps in more detailed career and contextual information. The enriched metadata will allow us to trace long-term developments such as gender balance, professionalization, regime legacies, and transnational entanglements, but they also expose structural biases and uneven documentation within both parliamentary and Wikidata sources. Future research can build on this linkage to connect speech content with prosopographical characteristics, investigate historical continuities and ruptures, and critically assess how political actors are represented in open knowledge graphs.

\printbibliography

\appendix

\section{Appendix / Tables}


{\footnotesize
\begin{longtable}{p{1.5cm}p{7cm}p{1.5cm}p{2cm}}
	\caption{Property Coverage Table}\label{tab:properties} \\
	\hline
	\textbf{Property} & \textbf{Property Label} & \textbf{n\_people} & \textbf{coverage\_percent} \\
	\hline
	\endfirsthead
	
	\hline
	\textbf{Property} & \textbf{Property Label} & \textbf{n\_people} & \textbf{coverage\_percent} \\
	\hline
	\endhead
	
	P21 & sex or gender & 4324 & 99.976879 \\
	P31 & instance of & 4324 & 99.976879 \\
	P569 & date of birth & 4323 & 99.953757 \\
	P1412 & languages spoken, written or signed & 4322 & 99.930636 \\
	P106 & occupation & 4320 & 99.884393 \\
	P27 & country of citizenship & 4309 & 99.630058 \\
	P19 & place of birth & 4303 & 99.491329 \\
	P227 & GND ID & 4297 & 99.352601 \\
	P735 & given name & 4297 & 99.352601 \\
	P214 & VIAF cluster ID & 4287 & 99.121387 \\
	P102 & member of political party & 4256 & 98.404624 \\
	P39 & position held & 4235 & 97.919075 \\
	P1559 & name in native language & 3997 & 92.416185 \\
	P528 & catalog code & 3862 & 89.294798 \\
	P937 & work location & 3830 & 88.554913 \\
	P103 & native language & 3739 & 86.450867 \\
	P734 & family name & 3529 & 81.595376 \\
	P1416 & affiliation & 3138 & 72.554913 \\
	P10632 & OpenSanctions ID & 2981 & 68.924855 \\
	P7902 & Deutsche Biographie (GND) ID & 2733 & 63.190751 \\
	P10832 & WorldCat Entities ID & 2491 & 57.595376 \\
	P373 & Commons category & 2375 & 54.913295 \\
	P18 & image & 2255 & 52.138728 \\
	P570 & date of death & 2204 & 50.959538 \\
	P646 & Freebase ID & 2180 & 50.404624 \\
	P1344 & participant in & 2166 & 50.080925 \\
	P13049 & DDB person (GND) ID & 2158 & 49.895954 \\
	P2671 & Google Knowledge Graph ID & 2132 & 49.294798 \\
	P213 & ISNI & 1909 & 44.138728 \\
	P1713 & biography at the Bundestag of Germany URL & 1884 & 43.560694 \\
	P20 & place of death & 1832 & 42.358382 \\
	P166 & award received & 1725 & 39.884393 \\
	P1284 & Munzinger person ID & 1646 & 38.057803 \\
	P5355 & abgeordnetenwatch.de politician ID & 1557 & 36.000000 \\
	P463 & member of & 1218 & 28.161850 \\
	P244 & Library of Congress authority ID & 1157 & 26.751445 \\
	P69 & educated at & 1037 & 23.976879 \\
	P856 & official website & 984 & 22.751445 \\
	P3368 & Prabook ID & 911 & 21.063584 \\
	P9964 & Kalliope-Verbund (GND) ID & 877 & 20.277457 \\
	P2013 & Facebook username & 858 & 19.838150 \\
	P3602 & candidacy in election & 813 & 18.797688 \\
	P2002 & X (Twitter) username & 804 & 18.589595 \\
	P1006 & Nationale Thesaurus voor Auteursnamen ID & 747 & 17.271676 \\
	P269 & IdRef ID & 725 & 16.763006 \\
	P13591 & Yale LUX ID & 707 & 16.346821 \\
	P1207 & NUKAT ID & 663 & 15.329480 \\
	P10553 & IxTheo authority ID & 574 & 13.271676 \\
	P140 & religion or worldview & 564 & 13.040462 \\
	P2003 & Instagram username & 543 & 12.554913 \\
	P4293 & PM20 folder ID & 497 & 11.491329 \\
	P345 & IMDb ID & 496 & 11.468208 \\
	P8168 & FactGrid item ID & 483 & 11.167630 \\
	P13983 & Cabinet minutes of the German Government ID & 461 & 10.658960 \\
	P9918 & Kallías ID & 450 & 10.404624 \\
	P8687 & social media followers & 449 & 10.381503 \\
	P1331 & PACE member ID & 430 & 9.942197 \\
	P108 & employer & 419 & 9.687861 \\
	P8189 & National Library of Israel J9U ID & 414 & 9.572254 \\
	P2397 & YouTube channel ID & 409 & 9.456647 \\
	P512 & academic degree & 401 & 9.271676 \\
	P691 & NL CR AUT ID & 368 & 8.508671 \\
	P11245 & YouTube handle & 365 & 8.439306 \\
	P6640 & JRC Names ID & 360 & 8.323699 \\
	P2163 & FAST ID & 358 & 8.277457 \\
	P2298 & NSDAP membership number (1925--1945) & 348 & 8.046243 \\
	P9885 & Bing entity ID & 344 & 7.953757 \\
	P7293 & National Library of Poland MMS ID & 318 & 7.352601 \\
	P13627 & Niedersächsische Personen ID & 306 & 7.075145 \\
	P5019 & Brockhaus Enzyklopädie online ID & 290 & 6.705202 \\
	P268 & Bibliothèque nationale de France ID & 289 & 6.682081 \\
	P26 & spouse & 265 & 6.127168 \\
	P119 & place of burial & 259 & 5.988439 \\
	P6634 & LinkedIn personal profile ID & 259 & 5.988439 \\
	P12458 & Parsifal cluster ID & 248 & 5.734104 \\
	P1442 & image of grave & 238 & 5.502890 \\
	P1971 & number of children & 232 & 5.364162 \\
	P13226 & Hessian Biography person (GND) ID & 224 & 5.179191 \\
	P1343 & described by source & 217 & 5.017341 \\
	P7085 & TikTok username & 214 & 4.947977 \\
	P1889 & different from & 209 & 4.832370 \\
	P1741 & GTAA ID & 202 & 4.670520 \\
	P1477 & birth name & 197 & 4.554913 \\
	P1695 & NLP ID & 188 & 4.346821 \\
	P40 & child & 178 & 4.115607 \\
	P648 & Open Library ID & 176 & 4.069364 \\
	P101 & field of work & 170 & 3.930636 \\
	P11892 & Threads username & 170 & 3.930636 \\
	P22 & father & 170 & 3.930636 \\
	P3065 & RERO ID (legacy) & 164 & 3.791908 \\
	P3987 & SHARE Catalogue author ID & 159 & 3.676301 \\
	P4985 & TMDB person ID & 156 & 3.606936 \\
	P10234 & Der Spiegel topic ID & 155 & 3.583815 \\
	P8445 & Our Campaigns candidate ID & 155 & 3.583815 \\
	P10 & video & 153 & 3.537572 \\
	P12361 & Bluesky handle & 153 & 3.537572 \\
	P11496 & CiNii Research ID & 151 & 3.491329 \\
	P271 & NACSIS-CAT author ID & 151 & 3.491329 \\
	P8748 & Rheinland-Pfälzische Personendatenbank (GND) ID & 149 & 3.445087 \\
	P1015 & BIBSYS ID & 146 & 3.375723 \\
	P1953 & Discogs artist ID & 138 & 3.190751 \\
	P2605 & ČSFD person ID & 135 & 3.121387 \\
	P12597 & museum-digital ID & 131 & 3.028902 \\
	P910 & topic's main category & 131 & 3.028902 \\
	P701 & Dodis ID & 128 & 2.959538 \\
	P866 & Perlentaucher ID & 127 & 2.936416 \\
	P551 & residence & 123 & 2.843931 \\
	P10916 & Süddeutsche Zeitung topic ID & 120 & 2.774566 \\
	P9671 & Bundesstiftung Aufarbeitung person ID & 115 & 2.658960 \\
	P1186 & MEP directory ID & 114 & 2.635838 \\
	P11249 & KBR person ID & 113 & 2.612717 \\
	P109 & signature & 107 & 2.473988 \\
	P2847 & Google+ ID & 104 & 2.404624 \\
	P2639 & Filmportal ID & 102 & 2.358382 \\
	P607 & participated in conflict & 102 & 2.358382 \\
	P7400 & LibraryThing author ID & 102 & 2.358382 \\
	P10608 & FID performing arts ID & 98 & 2.265896 \\
	P3267 & Flickr user ID & 95 & 2.196532 \\
	P2600 & Geni.com profile ID & 94 & 2.173410 \\
	P1280 & CONOR.SI ID & 93 & 2.150289 \\
	P1368 & National Library of Latvia ID & 90 & 2.080925 \\
	P6722 & FemBio ID & 90 & 2.080925 \\
	P396 & SBN author ID & 88 & 2.034682 \\
	P434 & MusicBrainz artist ID & 84 & 1.942197 \\
	P2949 & WikiTree person ID & 81 & 1.872832 \\
	P3373 & sibling & 79 & 1.826590 \\
	P4033 & Mastodon address & 79 & 1.826590 \\
	P13183 & JudaicaLink person (GND) ID & 76 & 1.757225 \\
	P949 & National Library of Israel ID (old) & 74 & 1.710983 \\
	P4342 & Great Norwegian Encyclopedia ID & 73 & 1.687861 \\
	P12412 & Lobbypedia ID & 72 & 1.664740 \\
	P7699 & National Library of Lithuania ID & 72 & 1.664740 \\
	P935 & Commons gallery & 72 & 1.664740 \\
	P25 & mother & 70 & 1.618497 \\
	P8313 & Lex ID & 70 & 1.618497 \\
	P1196 & manner of death & 67 & 1.549133 \\
	P509 & cause of death & 64 & 1.479769 \\
	P7305 & Online PWN Encyclopedia ID & 62 & 1.433526 \\
	P535 & Find a Grave memorial ID & 60 & 1.387283 \\
	P1185 & Rodovid ID & 59 & 1.364162 \\
	P5739 & Pontificia Università della Santa Croce ID & 54 & 1.248555 \\
	P9037 & BHCL UUID & 54 & 1.248555 \\
	P3222 & NE.se ID & 53 & 1.225434 \\
	P3430 & SNAC ARK ID & 53 & 1.225434 \\
	P6414 & DIZIE ID & 53 & 1.225434 \\
	P2604 & Kinopoisk person ID & 52 & 1.202312 \\
	P2372 & ODIS ID & 51 & 1.179191 \\
	P13429 & Saarland Biografien ID & 49 & 1.132948 \\
	P1038 & relative & 47 & 1.086705 \\
	P3847 & Open Library subject ID & 46 & 1.063584 \\
	P2190 & C-SPAN person string ID (deprecated) & 45 & 1.040462 \\
	P349 & NDL Authority ID & 45 & 1.040462 \\
	P1315 & NLA Trove people ID & 44 & 1.017341 \\
	P443 & pronunciation audio & 44 & 1.017341 \\
	\hline
\end{longtable}
}

	{\footnotesize 
	\begin{longtable}{p{1cm}p{3.5cm}p{3.5cm}p{1cm}p{1.5cm}}
		\caption{Statements Coverage Table}\label{tab:statements} \\
		\hline
		property & propertyLabel & objectDisplay & n\_people & coverage\_percent \\
		\hline
		\endfirsthead
		
		\hline
		property & propertyLabel & objectDisplay & n\_people & coverage\_percent \\
		\hline
		\endhead

		P31 & instance of & human (Q5) & 4324 & 99.976879 \\
		P1412 & languages spoken, written or signed & German (Q188) & 4313 & 99.722543  \\
		P27 & country of citizenship & Germany (Q183) & 4294 & 99.283237 \\
		P106 & occupation & politician (Q82955) & 4230 & 97.803468 \\
		P39 & position held & member of the German Bundestag (Q1939555) & 3785 & 87.514451 \\
		P103 & native language & German (Q188) & 3719 & 85.988439 \\
		P21 & sex or gender & male (Q6581097) & 3435 & 79.421965 \\
		P937 & work location & Bonn (Q586) & 2795 & 64.624277 \\
		P1416 & affiliation & CDU/CSU Bundestag fraction (Q1023134) & 1494 & 34.543353 \\
		P102 & member of political party & Social Democratic Party of Germany (Q49768) & 1470 & 33.988439 \\
		P102 & member of political party & Christian Democratic Union (Q49762) & 1414 & 32.693642 \\
		P1416 & affiliation & SPD Bundestag fraction (Q2207512) & 1272 & 29.410405 \\
		P937 & work location & Berlin (Q64) & 927 & 21.433526 \\
		P21 & sex or gender & female (Q6581072) & 889 & 20.554913 \\
		P1344 & participant in & 1999 German presidential election (Q451110) & 793 & 18.335260 \\
		P1344 & participant in & 2010 German presidential election (Q707063) & 735 & 16.994220 \\
		P1344 & participant in & 2009 German presidential election (Q707051) & 730 & 16.878613 \\
		P1344 & participant in & 2012 German presidential election (Q670933) & 729 & 16.855491 \\
		P1344 & participant in & 2004 German presidential election (Q314532) & 723 & 16.716763 \\
		P1344 & participant in & 2017 German presidential election (Q23558762) & 718 & 16.601156 \\
		P3602 & candidacy in election & 2013 German federal election (Q555931) & 523 & 12.092486 \\
		P102 & member of political party & Free Democratic Party (Q13124) & 512 & 11.838150 \\
		P1344 & participant in & 1949 West German presidential election (Q314561) & 429 & 9.919075 \\
		P166 & award received & Commander's Cross of the Order of Merit of the Federal Republic of Germany (Q10905276) & 414 & 9.572254 \\
		P1416 & affiliation & FDP Bundestag fraction (Q1387991) & 408 & 9.433526 \\
		P102 & member of political party & Nazi Party (Q7320) & 369 & 8.531792 \\
		P3602 & candidacy in election & 2017 German federal election (Q15062956) & 350 & 8.092486 \\
		P39 & position held & mayor (Q30185) & 331 & 7.653179 \\
		P102 & member of political party & Christian Social Union of Bavaria (Q49763) & 326 & 7.537572 \\
		P166 & award received & Officer's Cross of the Order of Merit of the Federal Republic of Germany (Q10905334) & 324 & 7.491329 \\
		P106 & occupation & lawyer (Q40348) & 311 & 7.190751 \\
		P39 & position held & substitute member of the Parliamentary Assembly of the Council of Europe (Q65494724) & 306 & 7.075145 \\
		P166 & award received & Bavarian Order of Merit (Q672787) & 278 & 6.427746 \\
		P106 & occupation & jurist (Q185351) & 272 & 6.289017 \\
		P166 & award received & Cross of the Order of Merit of the Federal Republic of Germany (Q10905380) & 267 & 6.173410 \\
		P102 & member of political party & Alliance '90/The Greens (Q49766) & 257 & 5.942197 \\
		P463 & member of & Parliamentary Assembly of the Council of Europe (Q939743) & 249 & 5.757225 \\
		P39 & position held & Member of the European Parliament (Q27169) & 245 & 5.664740 \\
		P27 & country of citizenship & German Democratic Republic (Q16957) & 241 & 5.572254 \\
		P937 & work location & City of Brussels (Q239) & 240 & 5.549133 \\
		P39 & position held & Parliamentary Secretary in Germany (Q19731005) & 238 & 5.502890 \\
		P937 & work location & Strasbourg (Q6602) & 238 & 5.502890 \\
		P39 & position held & Representative of the Parliamentary Assembly of the Council of Europe (Q65494714) & 228 & 5.271676 \\
		P106 & occupation & university teacher (Q1622272) & 215 & 4.971098 \\
		P140 & religion or worldview & Catholic Church (Q9592) & 208 & 4.809249 \\
		P166 & award received & Great Cross with Star and Sash of the Order of Merit of the Federal Republic of Germany (Q10905171) & 203 & 4.693642 \\
		P166 & award received & Knight Commander's Cross of the Order of Merit of the Federal Republic of Germany (Q10905235) & 201 & 4.647399 \\
		P1344 & participant in & list of participants in the coalition talks between the CDU/CSU and SPD in 2013 (Q15829617) & 197 & 4.554913 \\
		P512 & academic degree & doctorate (Q849697) & 192 & 4.439306 \\
		P19 & place of birth & Berlin (Q64) & 189 & 4.369942 \\
		P39 & position held & member of the Landtag of North Rhine-Westphalia (Q17781726) & 183 & 4.231214 \\
		P937 & work location & Düsseldorf (Q1718) & 182 & 4.208092 \\
		P463 & member of & European Union Parliamentary Group in the German Bundestag (Q1375174) & 164 & 3.791908 \\
		P102 & member of political party & Die Linke (Q49764) & 156 & 3.606936 \\
		P937 & work location & Munich (Q1726) & 155 & 3.583815 \\
		P39 & position held & member of the Abgeordnetenhaus of Berlin (Q18327335) & 147 & 3.398844 \\
		P39 & position held & member of the Landtag of Bavaria (Q17586301) & 139 & 3.213873 \\
		P937 & work location & Hanover (Q1715) & 132 & 3.052023 \\
		P39 & position held & member of the Volkskammer (Q18557729) & 127 & 2.936416 \\
		P106 & occupation & judge (Q16533) & 123 & 2.843931 \\
		P39 & position held & member of the Landtag of Lower Saxony (Q17521638) & 123 & 2.843931 \\
		P106 & occupation & journalist (Q1930187) & 122 & 2.820809 \\
		P1343 & described by source & Obálky knih (Q67311526) & 122 & 2.820809 \\
		P20 & place of death & Bonn (Q586) & 120 & 2.774566 \\
		P735 & given name & Hans (Q632842) & 120 & 2.774566 \\
		P101 & field of work & politics (Q7163) & 113 & 2.612717 \\
		P102 & member of political party & Party of Democratic Socialism (Q152554) & 111 & 2.566474 \\
		P102 & member of political party & Centre Party (Q157537) & 106 & 2.450867 \\
		P20 & place of death & Berlin (Q64) & 104 & 2.404624 \\
		P937 & work location & Stuttgart (Q1022) & 104 & 2.404624 \\
		P937 & work location & Wiesbaden (Q1721) & 103 & 2.381503 \\
		P140 & religion or worldview & Lutheran Churches (Q5415686) & 102 & 2.358382 \\
		P39 & position held & member of the Landtag of Hesse (Q17519166) & 101 & 2.335260 \\
		P166 & award received & Order of Merit of Baden-Württemberg (Q445673) & 100 & 2.312139 \\
		P166 & award received & Grand Cross 1st class of the Order of Merit of the Federal Republic of Germany (Q10905105) & 99 & 2.289017 \\
		P39 & position held & Berlin member of the Bundestag (Q821448) & 99 & 2.289017 \\
		P19 & place of birth & Hamburg (Q1055) & 98 & 2.265896 \\
		P69 & educated at & Ludwig-Maximilians-Universität München (Q55044) & 98 & 2.265896 \\
		P937 & work location & Hamburg (Q1055) & 95 & 2.196532 \\
		P166 & award received & Order of Merit of North Rhine-Westphalia (Q318770) & 94 & 2.173410 \\
		P735 & given name & Karl (Q15731830) & 94 & 2.173410 \\
		P102 & member of political party & Alternative for Germany (Q6721203) & 92 & 2.127168 \\
		P39 & position held & member of the Landtag of Baden-Württemberg (Q17481175) & 92 & 2.127168 \\
		P39 & position held & member of the Hamburg Parliament (Q19360355) & 92 & 2.127168 \\
		P19 & place of birth & Munich (Q1726) & 91 & 2.104046 \\
		P69 & educated at & University of Bonn (Q152171) & 88 & 2.034682 \\
		P937 & work location & Mainz (Q1720) & 88 & 2.034682 \\
		P1971 & number of children & 2 [integer] & 85 & 1.965318 \\
		P39 & position held & member of Landtag of Rhineland-Palatinate (Q18618563) & 83 & 1.919075 \\
		P512 & academic degree & Doctor of Laws (Q959320) & 83 & 1.919075 \\
		P140 & religion or worldview & Protestant church (Q346575) & 82 & 1.895954 \\
		P106 & occupation & economist (Q188094) & 81 & 1.872832 \\
		P39 & position held & Landrat (Q514725) & 79 & 1.826590 \\
		P735 & given name & Wolfgang (Q2589157) & 79 & 1.826590 \\
		P735 & given name & Peter (Q2793400) & 77 & 1.780347 \\
		P20 & place of death & Munich (Q1726) & 74 & 1.710983 \\
		P735 & given name & Hermann (Q1158570) & 74 & 1.710983 \\
		P463 & member of & Parlamentarische Linke (Q2052863) & 73 & 1.687861 \\
		P140 & religion or worldview & Lutheranism (Q75809) & 69 & 1.595376 \\
		P102 & member of political party & Socialist Unity Party of Germany (Q49750) & 67 & 1.549133 \\
		P69 & educated at & University of Hamburg (Q156725) & 65 & 1.502890 \\
		P735 & given name & Josef (Q15730712) & 65 & 1.502890 \\
		P140 & religion or worldview & Catholicism (Q1841) & 64 & 1.479769 \\
		P39 & position held & Senator of Berlin (Q28978682) & 64 & 1.479769 \\
		P1971 & number of children & 3 [integer] & 63 & 1.456647 \\
		P607 & participated in conflict & World War II (Q362) & 63 & 1.456647 \\
		P735 & given name & Klaus (Q15635230) & 62 & 1.433526 \\
		P937 & work location & Kiel (Q1707) & 62 & 1.433526 \\
		P106 & occupation & writer (Q36180) & 60 & 1.387283 \\
		P735 & given name & Walter (Q499249) & 60 & 1.387283 \\
		P1196 & manner of death & natural causes (Q3739104) & 59 & 1.364162 \\
		P39 & position held & member of the Landtag of Schleswig-Holstein (Q18130496) & 59 & 1.364162 \\
		P106 & occupation & trade unionist (Q15627169) & 58 & 1.341040 \\
		P735 & given name & Wilhelm (Q11027623) & 58 & 1.341040 \\
		P735 & given name & Heinrich (Q2018484) & 58 & 1.341040 \\
		P1971 & number of children & 1 [integer] & 57 & 1.317919 \\
		P102 & member of political party & Christian Democratic Union (GDR) (Q49754) & 56 & 1.294798 \\
		P20 & place of death & Hamburg (Q1055) & 54 & 1.248555 \\
		P39 & position held & President of the Bundesrat of Germany (Q363637) & 54 & 1.248555 \\
		P69 & educated at & University of Tübingen (Q153978) & 54 & 1.248555 \\
		P735 & given name & Fritz (Q1158596) & 53 & 1.225434 \\
		P69 & educated at & Freie Universität Berlin (Q153006) & 52 & 1.202312 \\
		P735 & given name & Georg (Q1985538) & 52 & 1.202312 \\
		P734 & family name & Müller (Q8157228) & 51 & 1.179191 \\
		P106 & occupation & farmer (Q131512) & 50 & 1.156069 \\
		P166 & award received & Wilhelm Leuschner Medal (Q2571514) & 49 & 1.132948 \\
		P19 & place of birth & Stuttgart (Q1022) & 49 & 1.132948 \\
		P735 & given name & Werner (Q676823) & 49 & 1.132948 \\
		P937 & work location & Dresden (Q1731) & 49 & 1.132948 \\
		P463 & member of & Netzwerk Berlin (Q1978977) & 48 & 1.109827 \\
		P735 & given name & Franz (Q4925932) & 48 & 1.109827 \\
		P735 & given name & Gerhard (Q7996169) & 48 & 1.109827 \\
		P106 & occupation & political scientist (Q1238570) & 47 & 1.086705 \\
		P69 & educated at & Humboldt-Universität zu Berlin (Q152087) & 47 & 1.086705 \\
		P102 & member of political party & German Party (Q674695) & 45 & 1.040462 \\
		P1412 & languages spoken, written or signed & English (Q1860) & 45 & 1.040462  \\
		P19 & place of birth & Essen (Q2066) & 45 & 1.040462 \\
		P19 & place of birth & Cologne (Q365) & 44 & 1.017341 \\
		P39 & position held & member of the Landtag of Saxony (Q17334379) & 44 & 1.017341 \\
		P39 & position held & secretary of state (Q736559) & 44 & 1.017341 \\
		P735 & given name & Michael (Q4927524) & 44 & 1.017341 \\
		P607 & participated in conflict & World War I (Q361) & 42 & 0.971098 \\
		P69 & educated at & University of Göttingen (Q152838) & 42 & 0.971098 \\
		P69 & educated at & University of Cologne (Q54096) & 42 & 0.971098 \\
		P735 & given name & Friedrich (Q14038597) & 42 & 0.971098 \\
		P735 & given name & Otto (Q18029644) & 42 & 0.971098 \\
		P463 & member of & Young Union (Q497594) & 41 & 0.947977 \\
		P463 & member of & Atlantik-Brücke (Q756504) & 41 & 0.947977 \\
	\end{longtable}
}

{\footnotesize
\begin{longtable}{lrr}
	\caption{Place of Birth Statements (property P19, label place of birth)}\label{tab:placeofbirth} \\
	\hline
	place & n\_people & coverage\_percent \\
	\hline
	\endfirsthead

	\hline
	
	\hline
	\endhead
	
	\hline
	\endfoot
	
	\hline
	\endlastfoot

Berlin (Q64) & 189 & 4.369942 \\
Hamburg (Q1055) & 98 & 2.265896 \\
Munich (Q1726) & 91 & 2.104046 \\
Stuttgart (Q1022) & 49 & 1.132948 \\
Essen (Q2066) & 45 & 1.040462 \\
Cologne (Q365) & 44 & 1.017341 \\
Hanover (Q1715) & 39 & 0.901734 \\
Leipzig (Q2079) & 38 & 0.878613 \\
Frankfurt (Q1794) & 34 & 0.786127 \\
Düsseldorf (Q1718) & 33 & 0.763006 \\
Wrocław (Q1799) & 29 & 0.670520 \\
Bremen (Q24879) & 29 & 0.670520 \\
Kiel (Q1707) & 27 & 0.624277 \\
Kassel (Q2865) & 27 & 0.624277 \\
Duisburg (Q2100) & 26 & 0.601156 \\
Bochum (Q2103) & 25 & 0.578035 \\
Münster (Q2742) & 25 & 0.578035 \\
Dortmund (Q1295) & 24 & 0.554913 \\
Dresden (Q1731) & 24 & 0.554913 \\
Nuremberg (Q2090) & 24 & 0.554913 \\
Heidelberg (Q2966) & 23 & 0.531792 \\
Freiburg im Breisgau (Q2833) & 21 & 0.485549 \\
Magdeburg (Q1733) & 20 & 0.462428 \\
Lübeck (Q2843) & 20 & 0.462428 \\
Mainz (Q1720) & 19 & 0.439306 \\
Karlsruhe (Q1040) & 18 & 0.416185 \\
Brunswick (Q2773) & 18 & 0.416185 \\
Osnabrück (Q2916) & 18 & 0.416185 \\
Saarbrücken (Q1724) & 17 & 0.393064 \\
Bonn (Q586) & 17 & 0.393064 \\
Wuppertal (Q2107) & 16 & 0.369942 \\
Augsburg (Q2749) & 16 & 0.369942 \\
Königsberg (Q4120832) & 16 & 0.369942 \\
Göttingen (Q3033) & 15 & 0.346821 \\
Oberhausen (Q2838) & 14 & 0.323699 \\
Würzburg (Q2999) & 14 & 0.323699 \\
Hildesheim (Q3185) & 14 & 0.323699 \\
Potsdam (Q1711) & 13 & 0.300578 \\
Chemnitz (Q2795) & 13 & 0.300578 \\
Rostock (Q2861) & 13 & 0.300578 \\
Ludwigshafen (Q2910) & 13 & 0.300578 \\
Darmstadt (Q2973) & 13 & 0.300578 \\
Aachen (Q1017) & 12 & 0.277457 \\
Erfurt (Q1729) & 12 & 0.277457 \\
Bielefeld (Q2112) & 12 & 0.277457 \\
Mönchengladbach (Q2758) & 12 & 0.277457 \\
Hagen (Q2871) & 12 & 0.277457 \\
Koblenz (Q3104) & 12 & 0.277457 \\
Siegen (Q3167) & 12 & 0.277457 \\
Szczecin (Q393) & 12 & 0.277457 \\
Barmen (Q153974) & 11 & 0.254335 \\
Wiesbaden (Q1721) & 11 & 0.254335 \\
Mannheim (Q2119) & 11 & 0.254335 \\
Gelsenkirchen (Q2765) & 11 & 0.254335 \\
Halle (Saale) (Q2814) & 11 & 0.254335 \\
Zwickau (Q3778) & 11 & 0.254335 \\
Marburg (Q3869) & 11 & 0.254335 \\
Giessen (Q3874) & 11 & 0.254335 \\
Krefeld (Q2805) & 10 & 0.231214 \\
Regensburg (Q2978) & 10 & 0.231214 \\
Kaiserslautern (Q3758) & 10 & 0.231214 \\
Gdańsk (Q1792) & 9 & 0.208092 \\
Mülheim an der Ruhr (Q2899) & 9 & 0.208092 \\
Oldenburg (Q2936) & 9 & 0.208092 \\
Gera (Q3750) & 9 & 0.208092 \\
Constance (Q3834) & 9 & 0.208092 \\
Celle (Q3933) & 9 & 0.208092 \\
Herford (Q3971) & 9 & 0.208092 \\
Bremerhaven (Q2706) & 8 & 0.184971 \\
Paderborn (Q2971) & 8 & 0.184971 \\
Trier (Q3138) & 8 & 0.184971 \\
Tübingen (Q3806) & 8 & 0.184971 \\
Neunkirchen (Q6880) & 8 & 0.184971 \\
Elberfeld (Q702259) & 8 & 0.184971 \\
Rheydt (Q702411) & 8 & 0.184971 \\
Bad Kreuznach (Q7047) & 8 & 0.184971 \\
Leer (Q15984) & 7 & 0.161850 \\
Hamm (Q2880) & 7 & 0.161850 \\
Herne (Q2904) & 7 & 0.161850 \\
Offenbach am Main (Q3042) & 7 & 0.161850 \\
Remscheid (Q3097) & 7 & 0.161850 \\
Jena (Q3150) & 7 & 0.161850 \\
Gütersloh (Q3771) & 7 & 0.161850 \\
Wilhelmshaven (Q3857) & 7 & 0.161850 \\
Castrop-Rauxel (Q3898) & 7 & 0.161850 \\
Weimar (Q3955) & 7 & 0.161850 \\
Fulda (Q3963) & 7 & 0.161850 \\
Neuwied (Q3967) & 7 & 0.161850 \\
Schwäbisch Gmünd (Q4037) & 7 & 0.161850 \\
Schweinfurt (Q4126) & 7 & 0.161850 \\
Völklingen (Q14876) & 6 & 0.138728 \\
Wernigerode (Q15982) & 6 & 0.138728 \\
Solingen (Q2942) & 6 & 0.138728 \\
Bottrop (Q3069) & 6 & 0.138728 \\
Neheim-Hüsten (Q314660) & 6 & 0.138728 \\
Marl (Q3813) & 6 & 0.138728 \\
Worms (Q3852) & 6 & 0.138728 \\
Lüdenscheid (Q3893) & 6 & 0.138728 \\
Bayreuth (Q3923) & 6 & 0.138728 \\
Brandenburg an der Havel (Q3931) & 6 & 0.138728 \\
Herten (Q3985) & 6 & 0.138728 \\
Frankfurt (Oder) (Q4024) & 6 & 0.138728 \\
Offenburg (Q4044) & 6 & 0.138728 \\
Wetzlar (Q4178) & 6 & 0.138728 \\
Passau (Q4190) & 6 & 0.138728 \\
Katowice (Q588) & 6 & 0.138728 \\
Heilbronn (Q715) & 6 & 0.138728 \\
\end{longtable}
}

{\footnotesize
\begin{table}[ht]\label{tab:languages}
	\centering
	\caption{Language properties with number of people and coverage percentage}
	\scriptsize
	\begin{tabular}{lllr r}
		\hline
		property & propertyLabel & language & n\_people & coverage\_percent \\
		\hline
		P1412 & languages spoken, written or signed & German (Q188)  & 4313 & 99.722543 \\
		P103  & native language                     & German (Q188)  & 3719 & 85.988439 \\
		P1412 & languages spoken, written or signed & English (Q1860) & 45  & 1.040462 \\
		P6886 & writing language                    & German (Q188)  & 19   & 0.439306 \\
		P103  & native language                     & Turkish (Q256) & 15   & 0.346821 \\
		P1412 & languages spoken, written or signed & Turkish (Q256) & 15   & 0.346821 \\
		P1412 & languages spoken, written or signed & French (Q150)  & 14   & 0.323699 \\
		P1412 & languages spoken, written or signed & Polish (Q809)  & 6    & 0.138728 \\
		P1412 & languages spoken, written or signed & Russian (Q7737) & 5   & 0.115607 \\
		\hline
	\end{tabular}
\end{table}
}

\end{document}